\documentclass[
 reprint,
 superscriptaddress,
 amsmath,amssymb,
 aps,
 pra,
]{revtex4-2}

\usepackage{mathtools}
\usepackage{amsmath}
\usepackage{amssymb}
\usepackage{amsfonts}
\usepackage{latexsym}
\usepackage{comment}
\usepackage{siunitx}
\usepackage{mhchem}
\usepackage{physics}
\usepackage{amsthm}
\usepackage{braket}
\usepackage{xcolor}
\usepackage{url}
\usepackage{graphicx}
\usepackage{hyperref}

\graphicspath{{./pic/}}

\begin{document}
\title{Connection of surface code patches
based on entanglement distillation\\ without postselection}

\author{Moeto Mishima}
\email{moetomishima@gmail.com}
\affiliation{Department of Applied Physics, Graduate School of Engineering, The University of Tokyo, 7-3-1 Hongo, Bunkyo-ku, Tokyo 113-8656, Japan}

\author{Toshihiko Sasaki}
\affiliation{Department of Applied Physics, Graduate School of Engineering, The University of Tokyo, 7-3-1 Hongo, Bunkyo-ku, Tokyo 113-8656, Japan}
\affiliation{Photon Science Center, Graduate School of Engineering, The University of Tokyo, 7-3-1 Hongo, Bunkyo-ku, Tokyo 113-8656, Japan}

\author{Yasunari Suzuki}
\affiliation{Computer and Data Science Laboratories, NTT, Inc., Musashino 180-8585, Japan} 
\affiliation{Center for Quantum Computing, RIKEN, 2-1 Hirosawa, Wako, Saitama 351-0198, Japan}

\author{Koji Azuma}
\affiliation{Basic Research Laboratories, NTT, Inc., 3-1 Morinosato Wakamiya, Atsugi, Kanagawa 243-0198, Japan}
\affiliation{NTT Research Center for Theoretical Quantum Information, NTT, Inc., 3-1 Morinosato Wakamiya, Atsugi, Kanagawa 243-0198, Japan}

\author{Masato Koashi}
\affiliation{Department of Applied Physics, Graduate School of Engineering, The University of Tokyo, 7-3-1 Hongo, Bunkyo-ku, Tokyo 113-8656, Japan}
\affiliation{Photon Science Center, Graduate School of Engineering, The University of Tokyo, 7-3-1 Hongo, Bunkyo-ku, Tokyo 113-8656, Japan}

\begin{abstract}
Distributed quantum computing is a promising approach for overcoming the scalability limitations of fault-tolerant quantum computing. Entanglement distillation, a process that extracts higher-fidelity entangled states from noisy ones, plays a key role in distributed quantum computing. However, simple distillation protocols typically rely on postselection, which poses practical challenges for implementation. Here, we propose a scheme for distributed surface-code quantum computing that benefits from entanglement distillation even without postselection. Numerical simulations show that the logical error rates in the proposed scheme are lower than those in the setting without distillation. Additionally, we demonstrate that the logical error rates can be suppressed further by exploiting the asymmetry between spacelike and timelike errors in the surface code. The proposed scheme offers a practical and simple approach to distributed quantum computing with surface codes.
\end{abstract}

\maketitle
\section{Introduction}
Fault-tolerant quantum computation plays a crucial role in enabling quantum computers to operate reliably while tolerating physical errors such as decoherence and gate errors. The surface code~\cite{kitaev2003fault, bravyi1998quantum, dennis2002topological, fowler2012surface} is one of the most promising quantum error-correcting codes because it can be implemented with a 2D array of qubits, has a high threshold~\cite{dennis2002topological, wang2003confinement, ohno2004phase, raussendorf2007topological, wang2011surface}, and supports efficient universal logical operations~\cite{horsman2012surface, fowler2018low, litinski2019game, chamberland2022universal}. Significant progress has been made in recent years in experimental demonstrations of the surface code~\cite{krinner2022realizing, google2023suppressing, acharya2024quantum}. 

A major drawback of surface codes is their low encoding rate, which makes it difficult to accommodate all the physical qubits required for large-scale computation on a single chip. To mitigate this problem and ensure scalability, we can use distributed quantum computing, where physical qubits are distributed across multiple nodes and internode operations are performed using shared physical Bell pairs~\cite{gottesman1998heisenberg, eisert2000optimal}. While several approaches to logical operations between nodes have been explored, remote lattice surgery~\cite{fowler2010surface,ramette2024fault} is one of the most promising approaches as it does not require postselection and allows a relatively high threshold of about 10\% for physical Bell pairs. For example, in superconducting-qubit architectures, remote lattice surgery can be implemented using a 1D array of microwave communication links along the edges of 2D qubit arrays. Achieving sufficiently low error rates compared to the 10\% threshold is still challenging for current microwave-link technologies. Therefore, several approaches perform entanglement distillation to generate clean physical Bell pairs from several noisy ones before consuming them in remote lattice surgery~\cite{marqversen2026fault,liu2026remote,leone2025resource}. This paper focuses on a protocol that uses entanglement distillation followed by remote lattice surgery.

There are two major approaches to entanglement distillation before remote lattice surgery: error-detecting distillation and error-correcting distillation. Error-detecting protocols detect errors in shared Bell pairs and determine whether to accept or reject trials. This approach can be implemented with shallow-depth circuits and enables compact distillation, as exemplified by the 2-to-1 protocol~\cite{bennett1996mixed, bennett1996purification}. However, this requires postselection. If distillation fails, it must be repeated until it succeeds, or a successful output pair must be supplied from elsewhere. These alternatives introduce additional waiting times or outcome-dependent routing and scheduling. The other approach, error-correcting distillation, estimates and corrects errors in shared Bell pairs based on measurement outcomes obtained during the distillation protocol. This approach avoids postselection, but protocols based on error-correcting codes, such as the $[[5,1,3]]$ perfect code, require more input Bell pairs and deeper circuits, which poses another challenge in device implementation. Therefore, protocol design has been forced into an unavoidable compromise: efficient error-detecting protocols with postselection or inefficient error-correcting protocols without postselection.

In this paper, we propose a scheme that breaks this trade-off: remote lattice surgery with error-detecting entanglement distillation without postselection. Specifically, we show that the 2-to-1 protocol, one of the simplest distillation protocols, can improve the logical error rates of remote lattice surgery even without postselection of successful cases. In our scheme, we use the original distillation circuit without modification, but retain the output state and use it to perform internode gates even when a rejection outcome is obtained. This enables a simple implementation of distributed quantum computing using surface codes by avoiding the long waiting times and complex scheduling associated with postselected entanglement distillation. A key reason for our logical-error-rate reduction without postselection is that the 2-to-1 distillation measurement outcomes provide soft information for error estimation in the following remote lattice surgery. More specifically, we consider two variants of the distillation protocol, one providing information about the bit-error ($X$-error) probability and the other about the phase-error ($Z$-error) probability of the output pair, and select between these two variants probabilistically. We incorporate this information into the surface-code decoder by providing the error probabilities conditioned on the measurement outcomes, enabling more accurate decoding.

We can further improve the performance of remote lattice surgery by exploiting the asymmetry between spacelike and timelike logical errors. Timelike logical errors can be suppressed by increasing the number of syndrome measurement rounds, which requires additional runtime but no additional spatial overhead. In the setting considered here, increasing the number of syndrome measurement rounds suppresses logical $Z$ errors. This makes it possible to increase the probability of choosing the distillation variant that provides information about $X$ errors, even though this variant increases the $Z$-error rate. By adjusting both the number of syndrome measurement rounds and the probability of choosing each distillation variant, we can further reduce the logical error rate at the cost of additional runtime.

We numerically evaluate the proposed scheme under a phenomenological noise model. Our simulations show that the logical error rates in the proposed scheme are lower than those without distillation. For example, when we consider only errors from the entangled states, the threshold against infidelity of the initial Bell pairs increases from $\SI{15}{\%}$ to $\SI{18}{\%}$. We then optimize the number of syndrome measurement rounds and the probability of choosing each distillation variant, and quantify the resulting additional reduction in logical error rates and the behavior of the optimal parameters.

\section{Proposed scheme}
We first describe the distributed surface-code architecture considered in this work and the existing distillation protocols on which our scheme is based. We then explain how we use the distillation outputs and measurement outcomes without postselection, and how we adjust the choice of distillation protocol together with the number of syndrome measurement rounds.

\subsection{Distributed surface-code architecture} \label{method_surface_code}

\begin{figure}[tbp]
\centering
  \includegraphics[width=0.5\linewidth]{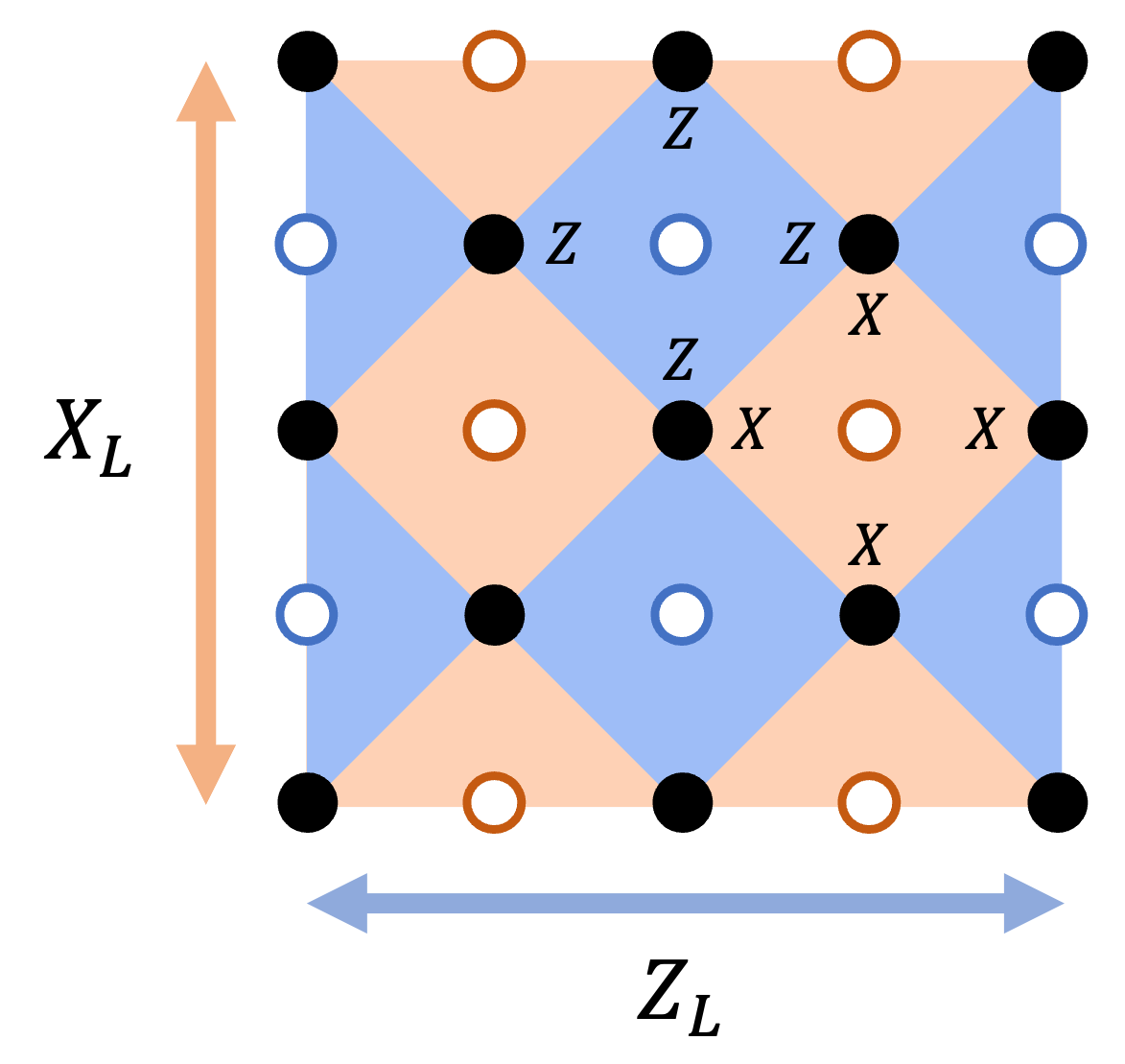}
  \caption{A patch of the planar surface code. Closed (resp.\ open) circles represent data (resp.\ ancilla) qubits. Orange (resp.\ blue) squares correspond to $X$ (resp.\ $Z$) stabilizer generators. The logical operator $X_{L}$ (resp.\ $Z_{L}$) corresponds to a vertical (resp.\ horizontal) line.}
  \label{surface_code_patch}
\end{figure}

The surface code is a 
Calderbank-Shor-Steane (CSS) stabilizer code defined on a two-dimensional square lattice. Data qubits and ancilla qubits are arranged alternately on the lattice, as shown in Fig.~\ref{surface_code_patch}. Stabilizer generators are defined as tensor products of Pauli $X$ or $Z$ operators acting on the nearest-neighbor data qubits of ancilla qubits. In Fig.~\ref{surface_code_patch}, orange (resp.\ blue) squares represent $X$ (resp.\ $Z$) stabilizers. Each stabilizer generator has weight at most four and can be measured using nearest-neighbor interactions between data and ancilla qubits. The logical operator $X_{L}$ (resp.\ $Z_{L}$) is defined as a tensor product of $X$ (resp.\ $Z$) operators acting on the data qubits along a path connecting the top and bottom (resp.\ left and right) boundaries. The code distance is given by the minimum weight of the logical operators, which corresponds to the vertical and horizontal lengths of the lattice. In a typical architecture of surface-code quantum computation, physical qubits are arranged on a 2D plane in a square lattice pattern enabling only nearest-neighbor interactions. Then logical qubits are defined using the parts of them as patches. Quantum computation is performed via patch deformation and lattice surgery which is achieved by modifying the set of measured stabilizers~\cite{litinski2019game}.

Following Ref.\ \cite{ramette2024fault}, we consider dividing the 2D plane into multiple smaller regions and implementing them as individual chips, as an approach to distributed quantum computing. We require that connectivity between physical qubits remains unchanged after the division of the plane. The CNOT gates corresponding to connections across chip boundaries are then implemented non-locally using entanglement.

\begin{figure}[tbp]
\centering
\includegraphics[width=0.85\linewidth]{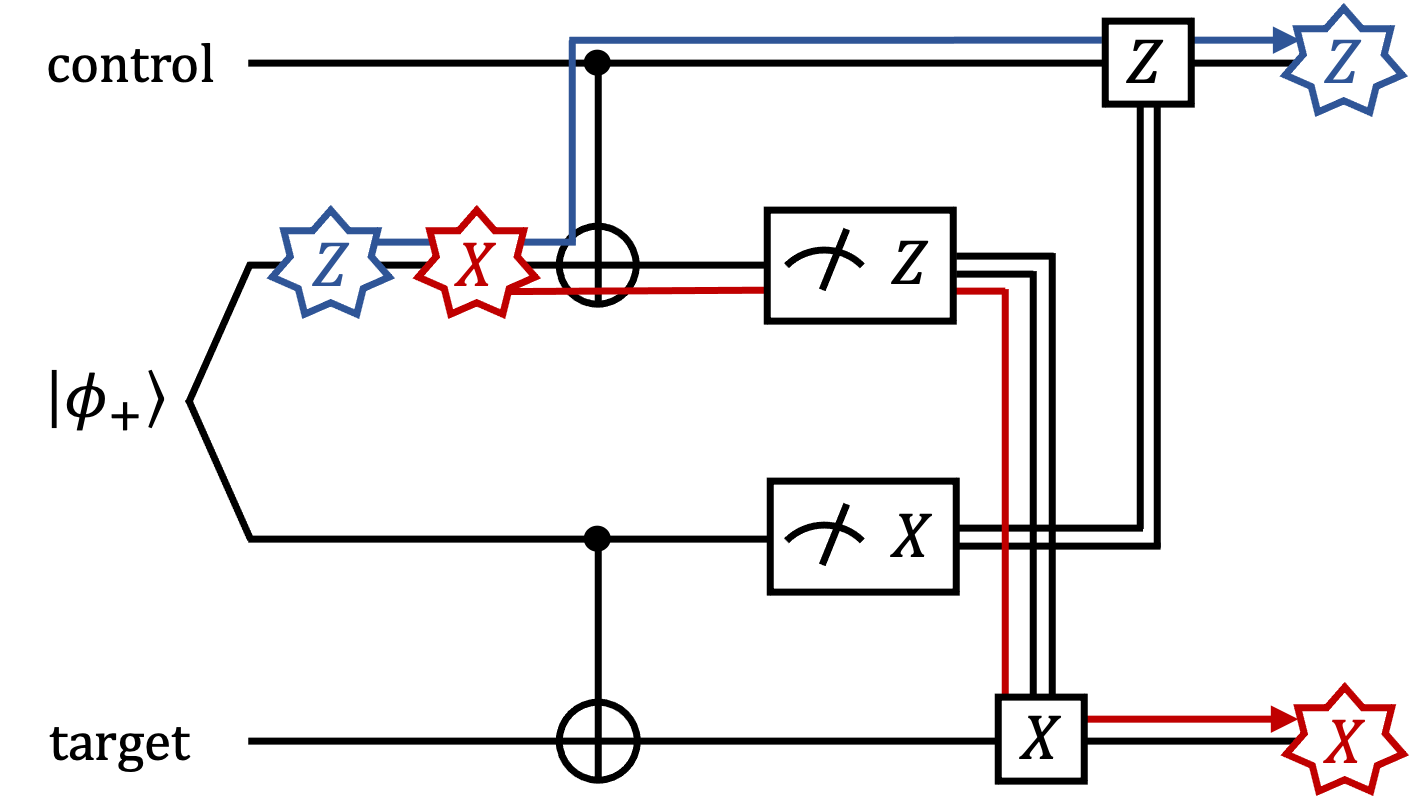}
  \caption{Non-local CNOT gate and propagation of Pauli errors. In this implementation, a $Z$ error propagates to the control qubit as a $Z$ error, while an $X$ error propagates to the target qubit as an $X$ error.}
  \label{nonlocal_cnot_error_propagation}
\end{figure}

The circuit for a non-local CNOT gate is shown in Fig.~\ref{nonlocal_cnot_error_propagation}. It is implemented using an entangled pair, local operations, and measurements of the two qubits that initially held the entangled pair~\cite{gottesman1998heisenberg,eisert2000optimal}. For later convenience, this figure also illustrates the propagation of Pauli errors through the non-local CNOT gate. As described in Sec.~\ref{subsec_distillation}, we consider noisy entangled pairs that can be regarded as ideal Bell pairs subjected to Pauli errors. A $Z$ error on the entangled pair propagates to the control qubit as a $Z$ error, while an $X$ error propagates to the target qubit as an $X$ error.

The non-local gates allow stabilizer generators to be measured across chip boundaries. These measurements enable lattice surgery between patches on different chips (remote lattice surgery). Depending on chip size, the same connectivity can also support a single code patch spanning multiple chips.

\subsection{Entanglement distillation} \label{subsec_distillation}

Shared Bell pairs can be used directly for remote lattice surgery~\cite{ramette2024fault}, or first processed by entanglement distillation to improve their fidelity. Ref.~\cite{liu2026remote} compares the resource costs of these two approaches. In each of its distillation-assisted settings, a fixed distillation protocol is used, and only successful output pairs are supplied to remote lattice surgery.

In this work, we focus on simple error-detecting distillation protocols that use two initial entangled pairs~\cite{bennett1996mixed, bennett1996purification}. We assume that the two input pairs are independent, that local operations and measurements in the distillation circuits are ideal, and that each input pair is an isotropic state with fidelity $F$
\begin{align}
\rho_{AB} = &  F\ket{\phi_+}\bra{\phi_+} \notag \\ & + \frac{1-F}{3}(\ket{\phi_-}\bra{\phi_-} + \ket{\psi_+}\bra{\psi_+} + \ket{\psi_-}\bra{\psi_-}),
\end{align}
where $\ket{\phi_\pm} = (\ket{00} \pm \ket{11})/\sqrt{2}$ and $\ket{\psi_\pm} = (\ket{01} \pm \ket{10})/\sqrt{2}$.

\begin{figure}[tbp]
\centering
  \includegraphics[width=0.85\linewidth]{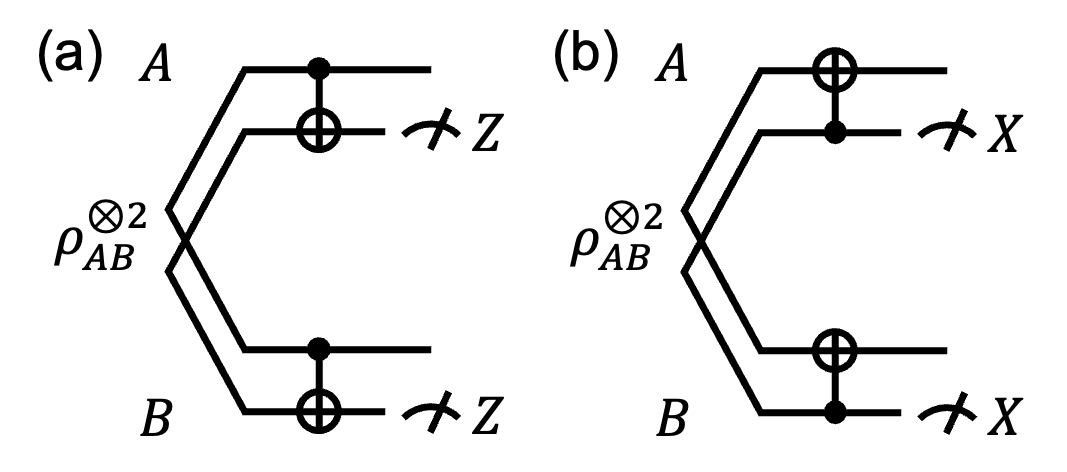}
  \caption{2-to-1 distillation protocols. (a) $X$ distillation for detecting $X$ errors and (b) $Z$ distillation for detecting $Z$ errors.}
  \label{distillation_circuit}
\end{figure}

Figure~\ref{distillation_circuit} shows the two variants considered here. When successful outputs are postselected, one reduces the $X$-error probability, while the other reduces the $Z$-error probability. We refer to these protocols as $X$ distillation and $Z$ distillation, respectively. The two variants are mutually convertible by Hadamard gates. Each protocol measures one of the input pairs and retains the other as its output. Agreement between the two measurement outcomes indicates success, while disagreement indicates failure. Under the assumptions above, a successful output has higher fidelity than the input for $1/2<F<1$, whereas a failed output is maximally mixed.

In conventional use, the outputs of these protocols are postselected: failed outputs are discarded, and the distillation process is repeated until it succeeds or a successful output is supplied from elsewhere if distillation fails. This introduces additional waiting times or outcome-dependent routing and scheduling. Error-correcting distillation can avoid postselection by using the measurement outcomes to estimate and correct errors. However, protocols based on error-correcting codes, such as the $[[5,1,3]]$ code~\cite{bennett1996mixed}, typically require more input Bell pairs and more complex circuits than the 2-to-1 protocols considered here.

\subsection{Proposed architecture}
\label{subsec_proposed_protocol}

\begin{table}[t]
    \centering
    \caption{Marginal error probabilities for $X$ distillation. The success and failure rows give probabilities conditioned on agreement and disagreement of the measurement outcomes, respectively. The total row gives their probability-weighted averages without postselection. The input row is included for comparison.}
    \label{tab:x-distillation-errors}
    \renewcommand{\arraystretch}{1.6}
    \begin{tabular}{lccc}
        \hline\hline
        Pair / outcome
        & Probability
        & $X$ error
        & $Z$ error \\
        \hline
        Input (raw)
        & ---
        & $\epsilon$
        & $\epsilon$ \\
        \hline
        Success
        & $P_{\mathrm{s}}$
        & $\epsilon^2/P_{\mathrm{s}}$
        & $\epsilon(1-\epsilon)/
                              P_{\mathrm{s}}$ \\[2pt]
        Failure
        & $1-P_{\mathrm{s}}$
        & $1/2$
        & $1/2$ \\
        \hline
        Total
        & $1$
        & $\epsilon$
        & $2\epsilon(1-\epsilon)$ \\
        \hline\hline
    \end{tabular}
\end{table}

\begin{figure*}[t]
  \includegraphics[width=\linewidth]{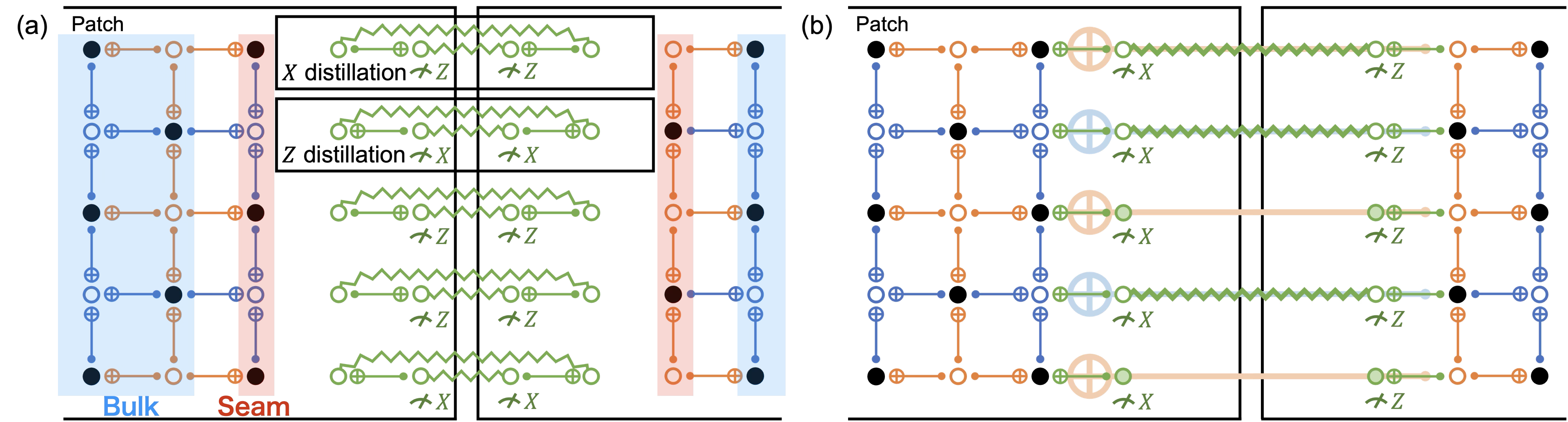}
  \caption{Implementation of the proposed protocol at a chip boundary. (a) $X$ or $Z$ distillation using two initial entangled pairs. (b) Non-local CNOT gates using the output pairs. The seam contains the surface-code qubits involved in these gates; the remaining surface-code qubits form the bulk. The choices of protocol, use of all outputs, and use of measurement outcomes for decoding are described in Sec.~\ref{subsec_proposed_protocol}.}
  \label{proposed_architecture}
\end{figure*}

Our scheme uses the two distillation protocols described above in the distributed surface-code architecture of Sec.~\ref{method_surface_code}. We retain every output pair and use it for a non-local CNOT gate, regardless of whether distillation succeeds or fails, without modifying the distillation circuits themselves. Although the average output fidelity is lower than the input fidelity, the measurement outcomes during the protocol provide information about the error probabilities of individual output pairs. We pass the corresponding conditional error probabilities to the surface-code decoder.

Table~\ref{tab:x-distillation-errors} summarizes the conditional and average error probabilities for $X$ distillation under the assumptions of Sec.~\ref{subsec_distillation}. We define $\epsilon=2(1-F)/3$ and the success probability $P_{\mathrm{s}}=(1-\epsilon)^2+\epsilon^2$. Here, an $X$ error denotes a Pauli $X$ or $Y$ error, and a $Z$ error denotes a Pauli $Z$ or $Y$ error. The probabilities for an individual output pair are conditioned on whether the two measurement outcomes agree or disagree; averaging these probabilities over both outcomes gives the error probabilities when the measurement outcomes are ignored.

For $0<\epsilon<1/2$, the $X$-error probability decreases when successful states are postselected. When successful and unsuccessful states are both retained, its average remains unchanged at $\epsilon$. In contrast, the $Z$-error probability increases even for successful outputs, and its average increases from $\epsilon$ to $2\epsilon(1-\epsilon)$. Thus, $X$ distillation provides information about the $X$-error probability of each output pair at the cost of an increase in the $Z$-error probability. The same argument applies to $Z$ distillation with $X$ and $Z$ interchanged. We use the probabilities conditioned on the distillation outcomes, rather than their averages, to set the decoder weights, as described in Sec.~\ref{subsec_noisemodel_decoding}.

Both $X$ and $Z$ errors must be suppressed in the surface code, but the two distillation variants provide different error information. We therefore select $X$ distillation with probability $q$ and $Z$ distillation with probability $1-q$. The parameter $q$ controls the relative amounts of information obtained about $X$ and $Z$ errors.

Figure~\ref{proposed_architecture} illustrates this scheme at the boundary of the chips. We refer to the surface-code qubits involved in non-local CNOT gates at the chip boundaries as seam qubits, and those interacting only with surface-code qubits within the same chip as bulk qubits, following Ref.~\cite{ramette2024fault}.

At each site and in each syndrome measurement round, two initial entangled pairs are supplied simultaneously. We independently select and perform one of the two distillation protocols according to $q$, and use its output for a non-local CNOT gate between seam qubits, using the circuit in Fig.~\ref{nonlocal_cnot_error_propagation}. These gates implement the stabilizer measurements across the seam. No output pair is discarded, and the success or failure of each distillation attempt is recorded for decoding.

In the implementation shown in Fig.~\ref{proposed_architecture}(a), we assume that entangled states are supplied independently at the inner and outer sites of the chips. Another possible implementation supplies entangled states only at the outer sites and subsequently swaps them to the inner sites. We use the former setting in the evaluation below.

We also treat the number $r$ of syndrome measurement rounds during remote lattice surgery as a parameter to be adjusted together with $q$. A conventional choice is $r=d$ for code distance $d$~\cite{horsman2012surface,liu2026remote}. For the evaluation task in Sec.~\ref{subsec_performance_evaluation_task}, which is a special case of remote lattice surgery, increasing $r$ suppresses the contribution of seam errors to the logical $Z$-error rate at the cost of increased runtime. When seam errors are dominant, this allows $X$ distillation to be selected more frequently, while the resulting increase in $Z$ errors is compensated for by increasing $r$. Additional rounds also accumulate bulk errors, so increasing $r$ is not always beneficial. We therefore optimize $q$ and $r$ jointly to reduce the logical error rate, as evaluated in Sec.~\ref{sec_numerical_results}.

With the assumed supply of two raw pairs per site and per round, the syndrome measurements can follow a fixed schedule without repetition or routing conditioned on distillation success. This is particularly useful when supplying substantially more entangled pairs per round is difficult. As shown in Sec.~\ref{sec_numerical_results}, our scheme reduces logical error rates compared with directly using raw entangled states, even at $r=d$ and $q=0.5$. This improvement requires no additional spatial resources beyond those needed to hold and process the two pairs used for distillation. Adjusting $r$ and $q$ can further reduce logical error rates. Increasing $r$ requires additional runtime but no additional spatial overhead.

\section{Evaluation method}
% \label{sec_evaluation_method}
In this section, we describe the methods used to evaluate the performance of the proposed scheme.

\subsection{Performance evaluation task}
\label{subsec_performance_evaluation_task}

\begin{figure}[t]
\centering
  \includegraphics[width=0.8\linewidth]{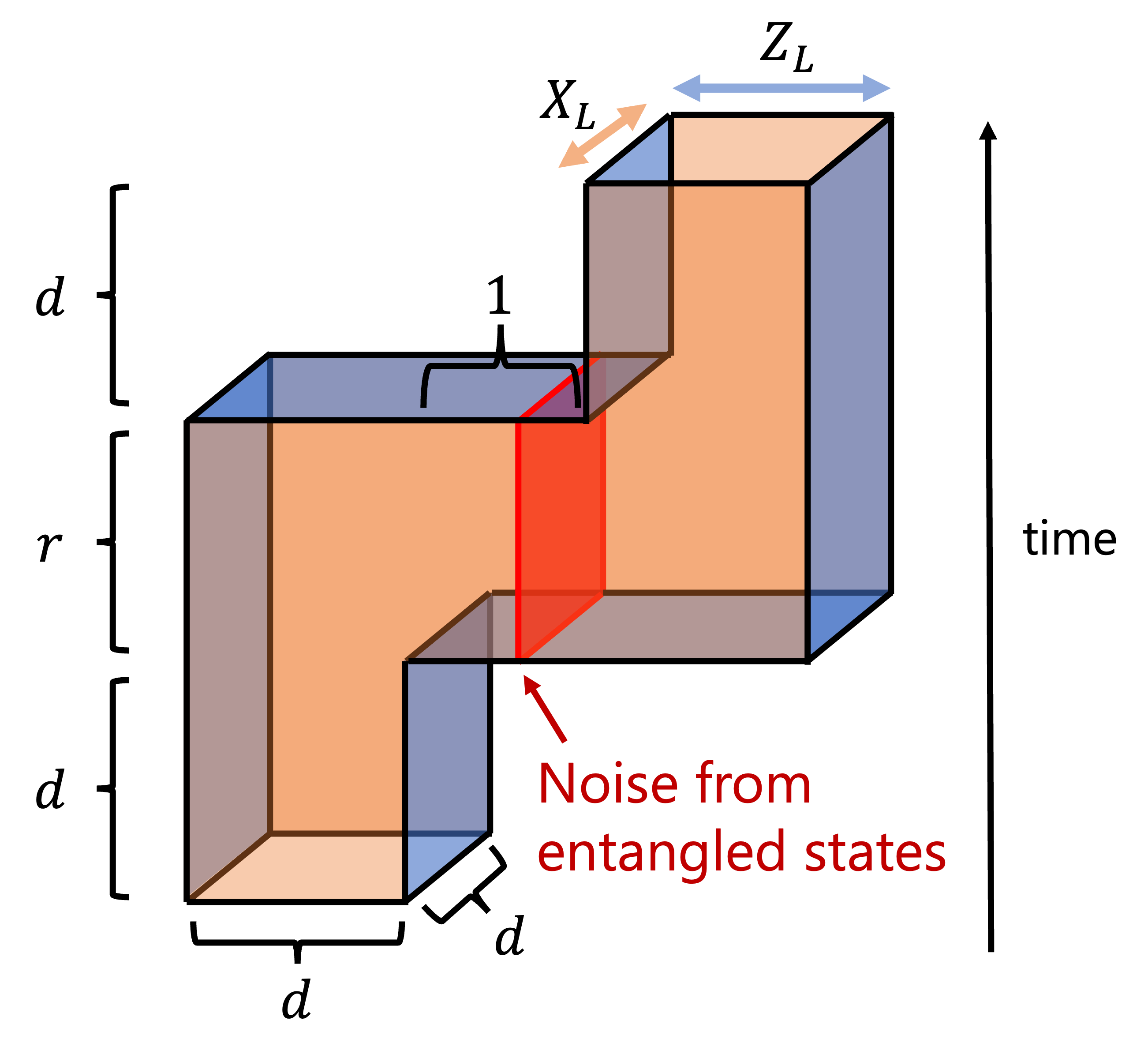}
  \caption{Decoding graph for moving a logical qubit across the seam. The decoding graphs for $X$ and $Z$ errors are overlaid. Orange and blue boundaries are open boundaries for $X$ and $Z$ errors, respectively. Error chains connecting boundaries of the same color correspond to logical errors. The error rates at the seam are higher than those in the bulk due to noisy entangled states.}
  \label{decoding_graph}
\end{figure}

To evaluate the performance of the proposed scheme, we consider a task of moving a logical qubit from one patch to another across the seam, which consists of a sequence of patch deformations, and calculate the logical error rate for this task.  Fig.~\ref{decoding_graph} illustrates the decoding graph in this task; see Sec.~\ref{subsec_noisemodel_decoding} for a detailed description of the decoding graph. First, a logical state is held on the left patch. Second, the code is extended to the right patch by initializing the data qubits in the right patch and those between the patches to $\ket{0}$, followed by measurements of the stabilizer generators of the extended code that consists of the qubits within and between both patches. Third, $r$ rounds of syndrome measurement are performed on the extended code to tolerate measurement errors, where $r$ is a tunable parameter. Increasing $r$ makes the logical state more robust to measurement errors. Fourth, the code is contracted to the right patch by measuring the data qubits in the left patch in the $Z$ basis. In this way, the logical state initially on the left patch is transferred to the right patch provided that no logical errors occur. We evaluate the logical $X$ and $Z$ error rates separately. To evaluate the logical $X$ (resp.\ $Z$) error rate, the logical qubit is initialized in $\ket{0_L}$ (resp.\ $\ket{+_L}$), and the logical $Z$ (resp.\ $X$) observable is measured after the logical qubit is moved. The logical error rate is defined as the probability that the logical measurement outcome is flipped from its ideal value.

In this setting, there is an asymmetry between logical $X$ and $Z$ errors at the seam. Increasing $r$ suppresses the contribution of seam errors to the logical $Z$-error rate. This asymmetry motivates the joint adjustment of $q$ and $r$ described in Sec.~\ref{subsec_proposed_protocol}. Its origin is explained in terms of the decoding graph in Sec.~\ref{subsec_noisemodel_decoding}.

The decoding graph in this task resembles that of lattice surgery~\cite{chamberland2022universal}. In fact, moving a logical qubit across the seam can be seen as a special case of remote lattice surgery, where an observable $X_{A}X_{B}$ is measured on the state $\ket{\psi_{A}} \ket{0_{B}}$, and the state on $A$ is subsequently measured in the $Z_{A}$ basis. Therefore, the logical error rates in lattice surgery are expected to behave similarly to those in this task.

Although it is possible to calculate the logical error rate in the simple memory experiment with a logical qubit defined across the chips, we chose not to adopt this method. When the seam is parallel to the support of a minimum-weight logical $X$ operator, the decoding problem lacks the timelike logical-error mechanism relevant to $Z$ errors~\cite{gidney2022stability}, leading to an underestimation of the logical $Z$ error rate. By considering the task of moving a logical qubit, which is sensitive to both logical $X$ and $Z$ error mechanisms, the impact of both $X$ and $Z$ errors will be evaluated properly.

\subsection{Noise model and decoding}
\label{subsec_noisemodel_decoding}
Here, we briefly review the decoding graph for the surface code; see Refs.~\cite{dennis2002topological,fujii2015quantum,demarti2024decoding} for a more detailed formulation. The decoding graph is a cubic lattice graph, where a horizontal edge represents a data qubit error and a vertical edge represents a syndrome measurement error. Each node is assigned the XOR of the current and previous stabilizer measurement results represented as binary values 0 and 1. This value is called a detector. In the absence of errors, all detectors have value zero. If an error occurs, the values of the detectors at the endpoints of the corresponding edge flip. Consequently, a detector has value 1 if errors occur on an odd number of its incident edges, and 0 if errors occur on an even number of its incident edges. In Fig.~\ref{decoding_graph}, the decoding graphs for $X$ and $Z$ errors, whose detectors are obtained from the $Z$- and $X$-stabilizer measurement outcomes, respectively, are overlaid. The orange and blue boundaries represent open boundaries on the decoding graph for $X$ and $Z$ errors, respectively, where each edge is connected to only one node. Decoding is the process of estimating which errors have occurred based on the stabilizer measurement results. In the decoding graph, this corresponds to finding a set of edges whose endpoints correspond to detectors with value 1. We refer to such a set of edges as an error chain. A logical error occurs if the symmetric difference between the actual error chain and the estimated error chain connects the open boundaries as a result of decoding. The asymmetry described in Sec.~\ref{subsec_performance_evaluation_task} can be understood from the decoding graph in Fig.~\ref{decoding_graph}. The shortest logical error chains at the seam have lengths $d$ for $X$ errors and $r$ for $Z$ errors, corresponding to spacelike and timelike logical errors, respectively. Therefore, increasing $r$ can exponentially suppress the contribution of seam errors to the logical $Z$-error rate.

We adopt a phenomenological noise model where each data qubit is subject to a depolarizing channel
\begin{align}
\mathcal{E}(\rho) = \qty(1-\frac{3}{2}p)\rho + \frac{p}{2}(X \rho X + Y \rho Y + Z \rho Z),
\end{align}
before each syndrome measurement, and measurement results flip with probability $p$. We refer to $p$ as the bulk error probability. In addition to data qubit errors and measurement errors, we consider errors originating from non-local CNOT gates due to noisy entangled states. As described in Sec.~\ref{subsec_distillation}, we consider only Bell diagonal states as noisy entangled states. As illustrated in Fig.~\ref{nonlocal_cnot_error_propagation}, when a non-local CNOT gate is performed via entangled states with a Pauli error, a $Z$ error propagates to the control qubit as a $Z$ error, and an $X$ error propagates to the target qubit as an $X$ error. These propagated errors commute with the non-local CNOT gate itself. If the propagated error acts on a surface-code ancilla qubit used for stabilizer measurement, it commutes with the other CNOT gates in the stabilizer measurement circuit and can be regarded as a measurement error. For an error propagated to a data qubit, in contrast, the set of detectors whose values are flipped generally depends on the CNOT ordering. In the CNOT schedule used here, the non-local CNOT gates at the seam are placed in the first or fourth layer of the four CNOT layers for stabilizer measurements. Since the propagated errors commute with the non-local CNOT gate itself, an error propagated to a data qubit can be regarded as occurring immediately before the first CNOT layer or immediately after the fourth CNOT layer. It therefore flips the same detectors as a data-qubit error in the phenomenological noise model. Consequently, the noise model considered here reduces to a phenomenological noise model where the data qubit error rates and the measurement error rates on the seam are higher than those in the bulk.

As described in Secs.~\ref{subsec_distillation} and~\ref{subsec_proposed_protocol}, our performance evaluation assumes that independent entangled pairs with the same fidelity $F$ are supplied at the inner and outer sites of the chip. In the alternative implementation, the additional swap operations can lead to different fidelities for the two input pairs. We leave simulations with unequal input fidelities for future work.

We use minimum weight perfect matching (MWPM), implemented in PyMatching 2~\cite{pymatchingv2}, as a decoder and decode $X$ and $Z$ errors separately. In MWPM decoding, a weight $w_i = \log((1-p_i)/p_i)$ is assigned to each edge based on its error probability $p_i$. The decoder outputs a set of edges with minimal total weight, subject to the constraint that it is consistent with the stabilizer measurement results. In other words, the output corresponds to the most probable error configuration given the stabilizer outcomes. As described in Sec.~\ref{subsec_proposed_protocol}, information on whether distillation has succeeded or failed at each site in each syndrome measurement round is incorporated into the decoder. The error probability $p_i$ of the corresponding edge, and hence its weight $w_i$, is adjusted according to whether distillation has succeeded or failed, enabling more accurate decoding.

\section{Numerical Results}
\label{sec_numerical_results}
In this section, we present the results of numerical simulations of the proposed scheme. We conducted simulations using Stim~\cite{gidney2021stim}. Each subsequent data point was obtained from $10^7$ Monte Carlo trials.

\subsection{Symmetric setting}
Typically, the number $r$ of rounds of a syndrome measurement on the extended patch is set to approximately $d$ when the error probability of data qubits and that of syndrome measurements are comparable, as in our noise model~\cite{horsman2012surface,liu2026remote}. Thus, we start with the setting $r=d$. When considering seam noise alone, the decoding problem is confined to the two-dimensional plane spanned by the spatial direction along the seam and the time direction. When $r=d$, the decoding graphs for $X$ and $Z$ errors are equivalent $d\times d$ lattices under the exchange of the spatial and temporal directions. We therefore consider $q=0.5$ for simplicity; in other words, $X$ and $Z$ distillations are selected with equal probability. We refer to this case as the symmetric setting. We denote the logical $X$ and $Z$ error rates by $p_{L,X}$ and $p_{L,Z}$, respectively, and use $p_L=\max{\{p_{L,X},p_{L,Z}\}}$ as the performance metric.

\begin{figure*}[t]
\centering
\includegraphics[width=2\columnwidth]{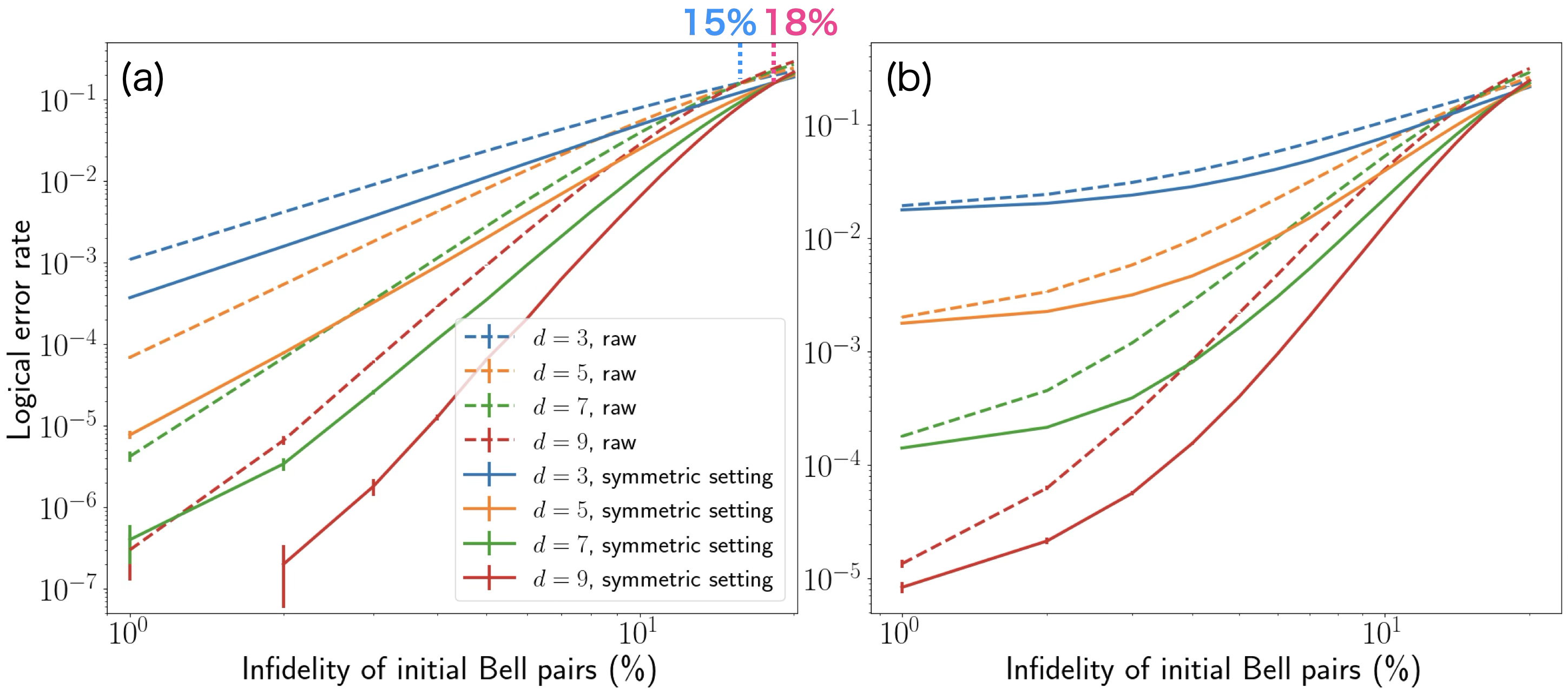}
    \caption{Comparison of logical error rates between cases where raw entangled pairs are used and where distillation is performed without postselection in the symmetric setting ($r=d$ and $q=0.5$). (a) Bulk error rate $p = \SI{0}{\%}$ and (b) $p = \SI{0.5}{\%}$. The logical error rate $p_L$ decreases for all code distances through entanglement distillation, even without postselection. The magnitude of this reduction increases with code distance $d$. Additionally, the threshold improves from $\SI{15}{\%}$ to $\SI{18}{\%}$ when $p = \SI{0}{\%}$. The threshold values when raw entangled pairs are used are consistent with the previous studies~\cite{li2016hierarchical,ramette2024fault}.}
    \label{distillation_effect}
\end{figure*}

Figure~\ref{distillation_effect} shows the logical error rate $p_{L}$ when the raw entangled states are used and when the output states of entanglement distillation are utilized without postselection. The results indicate that logical error rates can be suppressed and a higher threshold for the noise of entangled states can be achieved by applying entanglement distillation, even without postselection. This improvement can be understood from the mechanism described in Sec.~\ref{subsec_proposed_protocol}. Even without postselection, whether $X$ (resp.\ $Z$) distillation succeeds or fails provides information about whether an $X$ (resp.\ $Z$) error has occurred. On the other hand, distillation increases the error rate in the complementary basis. The results demonstrate that the benefit of using this error information exceeds the negative impact of the increased error rate in the complementary basis when the corresponding non-uniform error probabilities are incorporated into MWPM decoding. MWPM decoding that considers non-uniform error rates, as employed in this study, has been widely adopted in various studies~\cite{xu2019high,iolius2022performance,tiurev2023correcting}.

For all parameter values shown in Fig.~\ref{distillation_effect}, $p_{L,X}$ is larger than or equal to $p_{L,Z}$, whether raw entangled states are used or distillation is performed in the symmetric setting. Hence, $p_L=p_{L,X}$ throughout the figure. When $p = \SI{0}{\%}$, $p_{L,X}$ and $p_{L,Z}$ should be exactly equal due to the symmetry between the corresponding decoding graphs described above. When $p = \SI{0.5}{\%}$, $p_{L,X}$ and $p_{L,Z}$ exhibit similar dependence on the infidelity of the initial Bell pairs, with $p_{L,X}$ consistently larger than $p_{L,Z}$. This is because $p_{L,X}$ is larger than $p_{L,Z}$ when only bulk errors are present.

\subsection{Optimization of syndrome measurement rounds and distillation ratio}
\begin{figure*}[tbp]
\centering
\includegraphics[width=2\columnwidth]{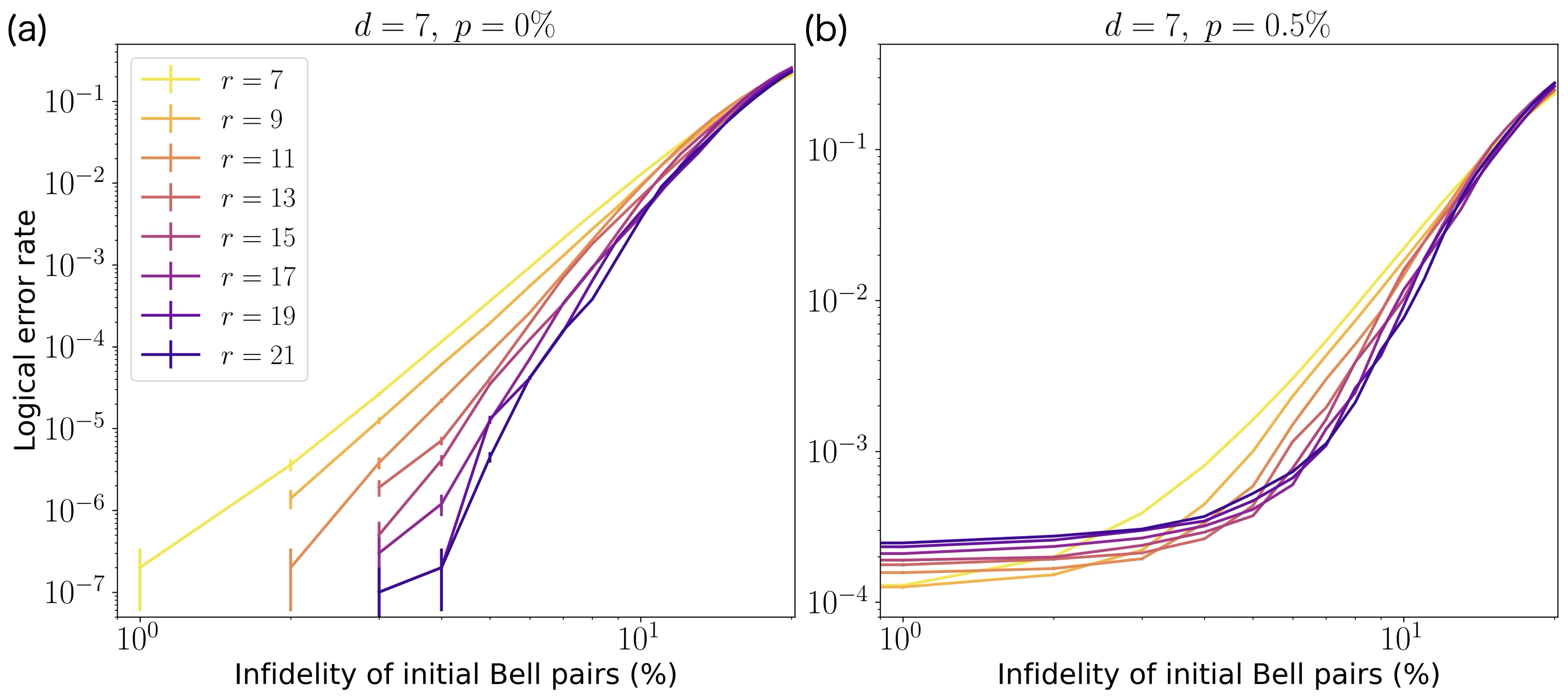}
    \caption{Logical error rate $p_{L} = \max{\{p_{L,X},p_{L,Z}\}}$ for several values of $r$, with $q$ optimized for each value of $r$, at $d=7$. (a) $p = \SI{0}{\%}$. $p_{L}$ decreases as $r$ increases over the simulated range. (b) $p = \SI{0.5}{\%}$. The optimal value $r_{\mathrm{opt}}$ is approximately $d$ when the infidelity of the initial Bell pairs is low and gradually increases with infidelity. For each data point, after optimizing $q$, the error bars of $p_{L,X}$ and $p_{L,Z}$ overlap, except when the optimum is attained at $q=1$. Thus, the condition $p_{L,X}\simeq p_{L,Z}$  expected at the optimum is satisfied within statistical uncertainty.}
    \label{logical_error_vs_r_q_optimized}
\end{figure*}

So far, we have adopted the conventional choice of $r=d$. In practice, $r$ can be set to a value larger than $d$ if additional runtime cost is accepted. In this case, the symmetry between the decoding graphs for $X$ and $Z$ errors at the seam is broken. As explained in Sec.~\ref{subsec_proposed_protocol}, increasing $r$ suppresses logical $Z$ errors at the seam, which allows $q$ to be increased to obtain more information about $X$ errors while compensating for the resulting increase in $Z$ errors. Figure~\ref{logical_error_vs_r_q_optimized} presents the logical error rate $p_{L}$ for several values of $r$, with $q$ optimized for each value of $r$, at $d=7$.

The results indicate that when $p = \SI{0}{\%}$, that is, when bulk errors are negligible, $p_{L}$ decreases as $r$ increases. In contrast, when $p = \SI{0.5}{\%}$, in other words, when bulk errors cannot be neglected, $p_L$ takes the minimum value at $r\simeq d$ in the low-infidelity regime. As the infidelity increases, however, larger values of $r$ lead to lower values of $p_L$. Let $r_{\mathrm{opt}}$ denote a value of $r$ at which $p_L$ attains its minimum.

\begin{figure*}[tbp]
\centering
\includegraphics[width=2\columnwidth]{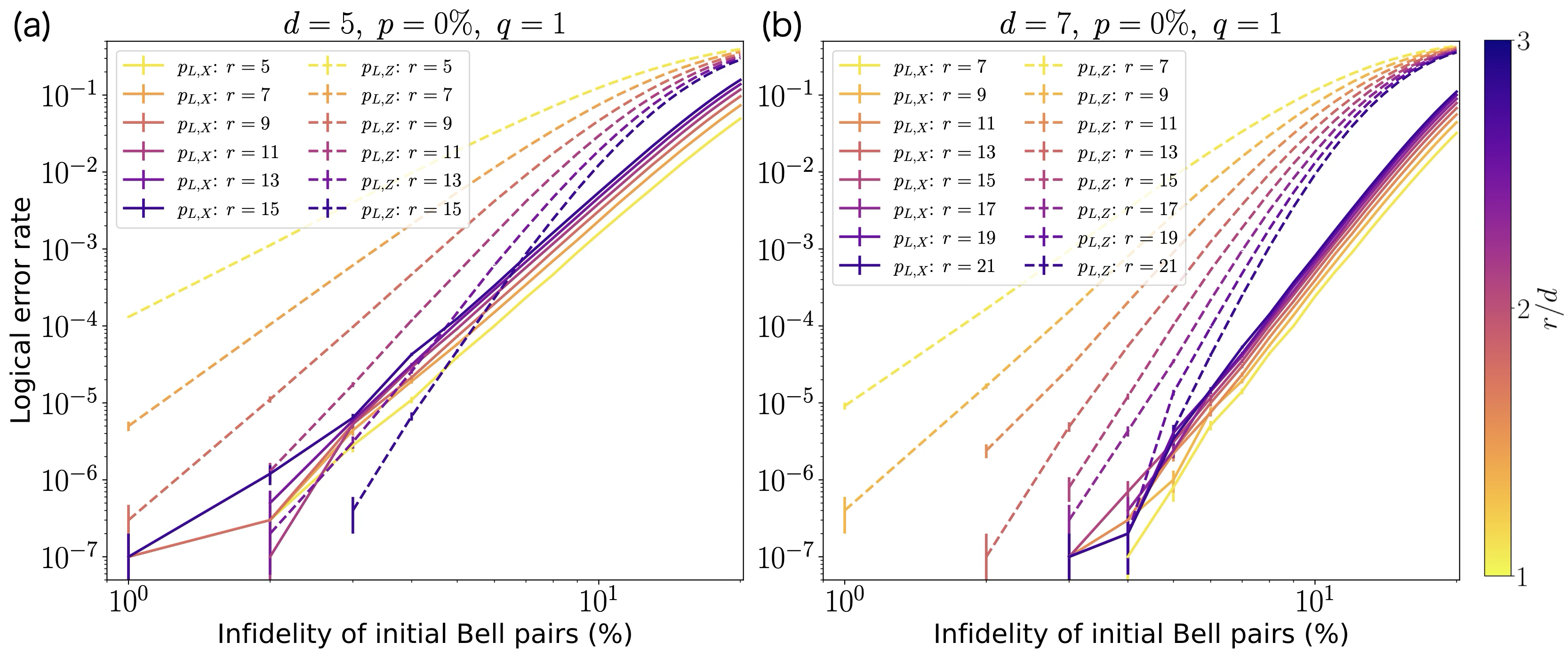}
    \caption{Logical $X$ error rate $p_{L,X}$ (solid lines) and logical $Z$ error rate $p_{L,Z}$ (dashed lines) as functions of the infidelity of the initial Bell pairs for several values of $r$, with $p = \SI{0}{\%}$ and $q = 1$. (a) $d=5$. (b) $d=7$. As $r$ increases, $p_{L,Z}$ decreases rapidly, while $p_{L,X}$ increases gradually. In addition, as the infidelity increases, the suppression of $p_{L,Z}$ with increasing $r$ becomes weaker, leading to larger $r_{\mathrm{opt}}$.}
    \label{logical_xz_error_vs_r}
\end{figure*}

The behavior for $p=\SI{0}{\%}$ can be understood as follows. As explained in Sec.~\ref{subsec_proposed_protocol}, increasing $r$ allows $q$ to be increased while compensating for the resulting increase in $Z$ errors. Once $q$ reaches its upper bound $q=1$, no additional information about $X$ errors can be obtained by further adjusting $q$. The minimum value of $p_L$ is then expected to be obtained at $q=1$ and a finite value of $r$. At first sight, however, Fig.~\ref{logical_error_vs_r_q_optimized} appears to show that the optimized logical error rate continues to decrease as $r$ increases over the entire range of infidelities considered. This does not imply that $r_{\mathrm{opt}}$ is infinite, because the simulations were restricted to $r\leq 3d$, and the optimum may lie beyond this range. To investigate this possibility and examine the mechanism that determines $r_{\mathrm{opt}}$, Fig.~\ref{logical_xz_error_vs_r} shows $p_{L,X}$ and $p_{L,Z}$ separately at $q=1$ for $d=5$ and $d=7$. When $r$ is small, $p_{L,Z}$ dominates. As $r$ increases, $p_{L,Z}$ decreases rapidly, while $p_{L,X}$ increases gradually. These trends are consistent with the expected exponential suppression of $p_{L,Z}$ and the approximately linear accumulation of $p_{L,X}$ as the number of rounds of syndrome measurement increases. The value of $r$ at which $p_{L,X}$ and $p_{L,Z}$ become comparable is therefore expected to be close to $r_{\mathrm{opt}}$. For smaller $r$, where $p_{L,Z}>p_{L,X}$ at $q=1$, decreasing $q$ can further reduce $p_L$. Beyond the crossover, further increase in $r$ is expected to result in an increase in $p_L$, suggesting the existence of a finite $r_{\mathrm{opt}}$. At $r=3d$, $p_{L,Z}$ remains larger than $p_{L,X}$ for the infidelities above approximately $\SI{7}{\%}$ for $d=5$ and $\SI{5}{\%}$ for $d=7$. In these regimes, the crossover has not been reached within the simulated range, suggesting that $r_{\mathrm{opt}}$ may exceed $3d$.

\begin{figure}[tbp]
\centering
\includegraphics[width=\columnwidth]{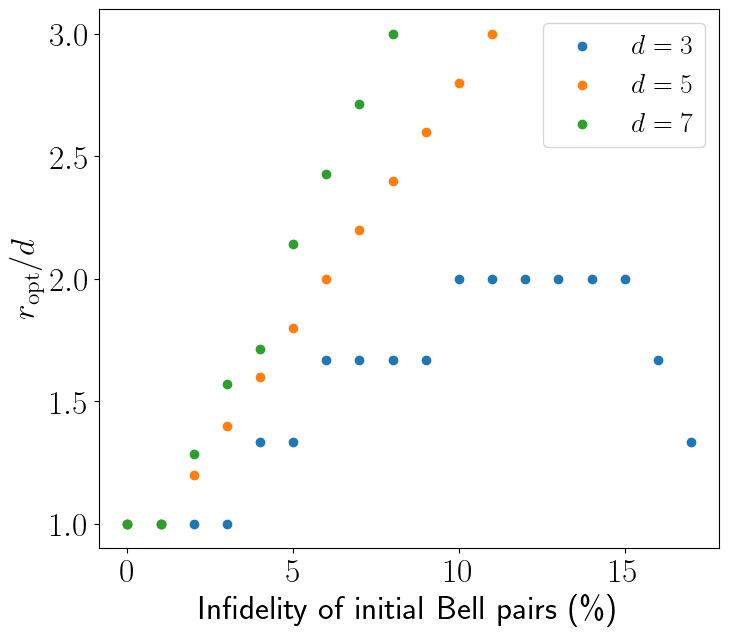}
    \caption{Ratio $r_{\mathrm{opt}}/d$ as a function of the infidelity of the initial Bell pairs for $p = \SI{0.5}{\%}$ and $d=3,5,7$. Only data points for which the minimum is attained within the simulated range $r\leq3d$ are plotted.}
    \label{r_opt_over_d}
\end{figure}

When $p = \SI{0.5}{\%}$, the accumulation of bulk errors over rounds has a considerable impact. In the range of low infidelity of entangled states, where bulk errors are dominant, increasing $r$ leads to accumulation of the logical $X$ errors in the bulk, resulting in higher logical error rates. Therefore, $r$ should be set to approximately $d$ in this range. As the infidelity increases gradually, the impact of the seam becomes larger, causing $r_{\mathrm{opt}}$ to increase progressively. Fig.~\ref{r_opt_over_d} presents the ratio $r_{\mathrm{opt}}/d$ as a function of the infidelity of the initial Bell pairs when $p = \SI{0.5}{\%}$. It shows that $r_{\mathrm{opt}}$ is approximately $d$ when the infidelity is small, and $r_{\mathrm{opt}}$ increases as the infidelity increases for any code distance $d$. In addition, $r_{\mathrm{opt}}/d$ increases as $d$ increases. When $d = 3$, $r_{\mathrm{opt}}$ decreases as the infidelity increases where the infidelity is near the threshold. In this region, the degree of decrease in $p_{L,Z}$ is extremely small due to the large infidelity, so increasing $r$ is not necessarily beneficial. In fact, $p_{L}$ hardly changes by adjusting $r$ and $q$ when the infidelity is near the threshold.

\begin{figure*}[htbp]
\centering
\includegraphics[width=2\columnwidth]{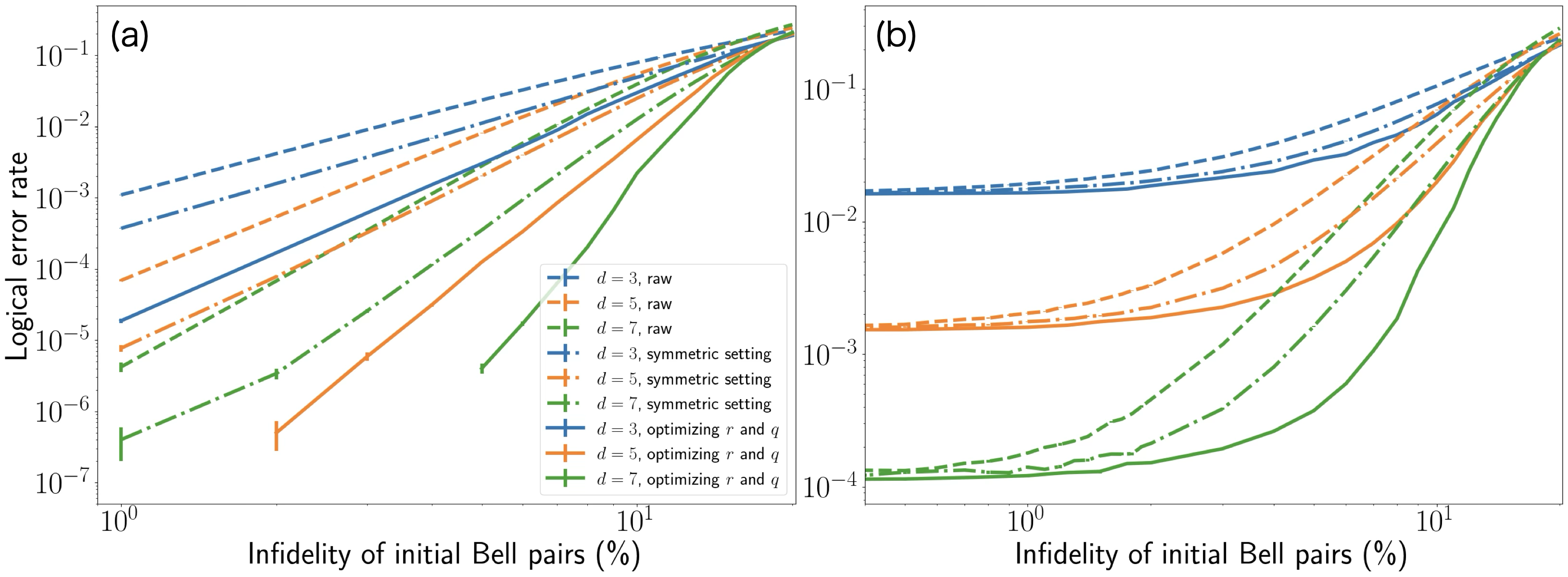}
    \caption{Logical error rates for three cases: when raw entangled pairs are used (dashed lines), when distillation is performed in the symmetric setting with $r=d$ and $q=0.5$ (dash-dotted lines), and when $q$ is optimized over its full range and $r$ is optimized over $r\leq 3d$ (solid lines). (a) $p=\SI{0}{\%}$. (b) $p=\SI{0.5}{\%}$.}
    \label{raw_symmetric_optimized_comparison}
\end{figure*}

Fig.~\ref{raw_symmetric_optimized_comparison} presents a comparison of logical error rates for three cases: (i) when raw Bell pairs are used, (ii) when distillation without postselection is performed in the symmetric setting, i.e., $r=d$ and $q=0.5$, and (iii) when distillation without postselection is performed with optimization over the full range of $q$ and over the range $r\leq 3d$.

The logical error rates are suppressed by introducing distillation and further by optimizing $r$ and $q$. When $p = \SI{0}{\%}$, both effects contribute to reducing the logical error rate over the entire range of infidelity considered. In contrast, when $p = \SI{0.5}{\%}$, introducing distillation alone is sufficient to mitigate the impact of seam errors for infidelities below approximately $\SI{2}{\%}$, and further adjustment of $r$ and $q$ has little additional effect on the logical error rate. For $d=5$ and $d=7$, however, optimizing $r$ and $q$ provides an additional reduction in the logical error rate over a broad range of infidelities above approximately $\SI{2}{\%}$ and below the threshold. It can be observed that optimizing $r$ and $q$ hardly changes the threshold for both $p = \SI{0}{\%}$ and $p = \SI{0.5}{\%}$.

The choice of the benchmark task is important when optimizing the relative amounts of $X$ and $Z$ distillation. As discussed in Sec.~\ref{subsec_performance_evaluation_task}, simply simulating memory experiments with a logical qubit defined across the chips leads to an underestimation of the impact of $Z$ errors. Indeed, in the task of moving a logical qubit across the seam, the numerical results show that both $p_{L,X}$ and $p_{L,Z}$ play important roles in determining the optimized logical error rate.

\section{Conclusion}
We have proposed and analyzed a scheme for distributed quantum computing with surface codes based on entanglement distillation without postselection, thereby avoiding the complexity associated with its implementation. We find that the use of simple 2-to-1 distillation protocols reduces the logical error rates relative to the case without distillation, even when the output states are used regardless of whether distillation succeeds or fails. When seam errors have a non-negligible impact compared to bulk errors, the logical error rate can be further reduced by increasing the number of syndrome measurement rounds, at the expense of additional runtime. In this regime, the ratio of $X$ and $Z$ distillation can be adjusted accordingly to further reduce the logical error rate.

The optimal number of syndrome measurement rounds and the appropriate distillation ratio are determined by the balance between logical $X$ and $Z$ errors. In the regime where seam errors are dominant, increasing the number of syndrome measurement rounds suppresses logical $Z$ errors, while increasing the ratio of $X$ distillation to $Z$ distillation helps suppress logical $X$ errors. In contrast, in the regime where bulk errors are dominant, increasing the number of syndrome measurement rounds leads to the accumulation of bulk errors. When both bulk and seam errors are relevant, these competing effects determine the optimal number of rounds and distillation ratio.

In this work, we employed a simple phenomenological noise model to evaluate logical error rates. Conducting numerical simulations under circuit-level noise would be an interesting direction for future research. Our analysis is also limited to some of the simplest distillation protocols. Investigating more complex distillation protocols involving more than two pairs remains an intriguing question~\cite{fujii2009entanglement, krastanov2019optimized, addala2023faster, jansen2022enumerating}.

\section*{Note Added}
Recently, we became aware of related work by Schmidt et al. \cite{schmidt2026resource}, which was developed independently of the present study. Their work provides a general framework for integrating entanglement distillation and quantum error correction in distributed quantum computing. The present work focuses on a minimal implementation using 2-to-1 distillation without postselection, incorporating the distillation outcomes into matching-based decoding through conditional error probabilities. We also jointly optimize the distillation-basis selection probability and the number of syndrome measurement rounds, and characterize the resulting performance improvements.

\section*{Acknowledgments}
We thank Yuuki Tokunaga and Shu Watanabe for valuable advice and discussions. 
This work is supported by MEXT Q-LEAP Grant No.~JPMXS0120319794, JPMXS0118068682, JST Moonshot R\&D Grant No.~JPMJMS256E, JST CREST Grant No.~JPMJCR23I4, JPMJCR24I4, JPMJCR25I4, MEXT Feasibility Study on the future HPCI.

\bibliography{biblio}

\end{document}